\documentclass[aps,prl,superscriptaddress,twocolumn,showpacs]{revtex4-1}

\usepackage{amsmath, amsthm, amssymb}
\usepackage{graphicx}
\usepackage{subfigure}
\usepackage{color}
\usepackage{times}

\usepackage{natbib}
\usepackage{hyperref}
\hypersetup{
        colorlinks=true,
}

\newcommand{\eqnref}[1]{(\ref{#1})}

\newcommand{\be}{\begin{equation} }
\newcommand{\ee}{\end{equation} }
\newcommand{\ba}{\begin{eqnarray} }
\newcommand{\ea}{\end{eqnarray} }

\renewcommand{\vec}[1]{\mathbf{ #1 }}

\begin{document}

\title{Helical order in one-dimensional magnetic atom chains and possible emergence of \\ Majorana bound states}

\author{Younghyun Kim}
\affiliation{Physics Department, University of California,  Santa Barbara, CA 93106, USA}

\author{Meng Cheng}
\affiliation{Station Q, Microsoft Research, Santa Barbara, CA 93106-6105, USA}

\author{Bela Bauer}
\affiliation{Station Q, Microsoft Research, Santa Barbara, CA 93106-6105, USA}

\author{Roman M. Lutchyn}
\affiliation{Station Q, Microsoft Research, Santa Barbara, CA 93106-6105, USA}

\author{S. Das Sarma}
\affiliation{Condensed Matter Theory Center and Joint Quantum Institute, Department of Physics, University of Maryland, College Park, MD 20742, USA}

\begin{abstract}
We theoretically obtain the phase diagram of localized magnetic impurity spins arranged in a
one-dimensional chain on top of a one- or two-dimensional electron gas. The interactions
between the spins are mediated by the Ruderman-Kittel-Kasuya-Yosida (RKKY) mechanism
through the electron gas. Recent work predicts that such a system may intrinsically support topological
superconductivity without spin-orbit coupling when a helical spin-density wave is spontaneously
formed in the spins, and superconductivity is induced in the electron gas.
We analyze, using both analytical and numerical techniques, the conditions under which
such a helical spin state is stable in a realistic situation in the presence of disorder. We show that
(i) it appears only when the spins are coupled to a (quasi-) 1D electron gas, and
(ii) it becomes unstable towards the formation of (anti) ferromagnetic domains if the disorder in the impurity spin positions $\delta R$ becomes comparable with the Fermi wave length.
We also examine the stability of the helical state against Gaussian potential disorder
in the electronic system using a diagrammatic approach. Our results suggest that in order to stabilize the
helical spin state and thus the emergent topological superconductivity under realistic experimental conditions,
a sufficiently strong Rashba spin-orbit coupling, giving rise to Dzyaloshinskii-Moriya interactions, is required.
\end{abstract}


\maketitle

\begin{figure}
  \includegraphics[width=7.4cm]{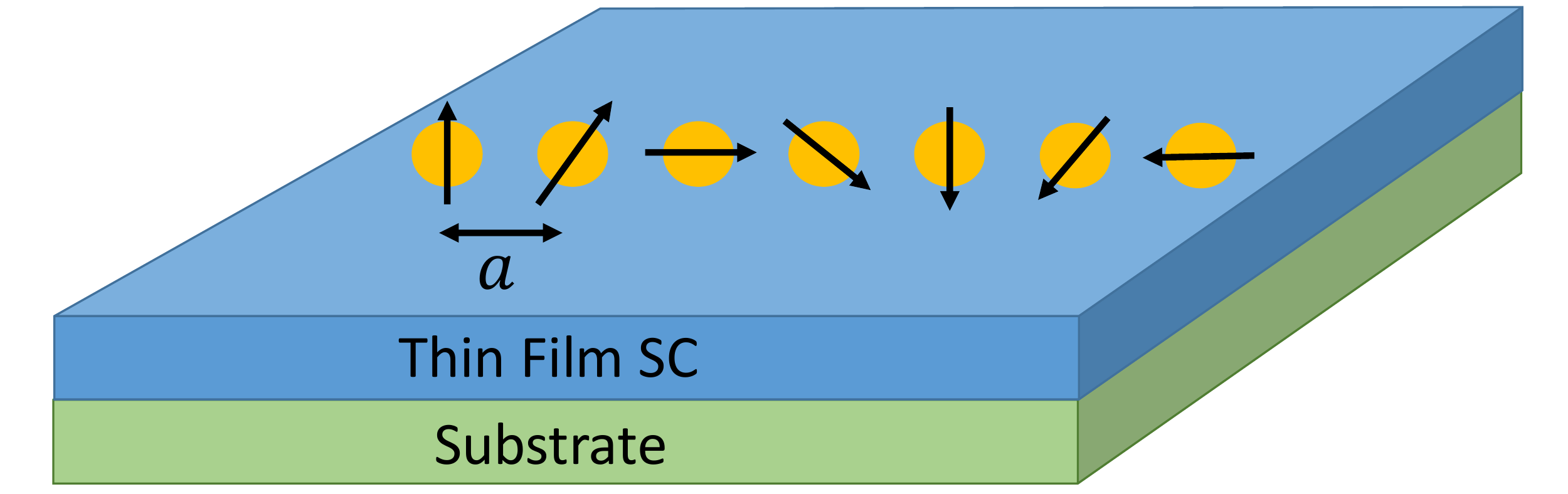}
        \caption{(Color online) Schematic plot of the experimental setup: magnetic impurity atoms are placed on top of a thin film superconductor with strong Rashba spin-orbit interaction, e.g., ultrathin Pb film on Si(111) substrate~\cite{Upton04, Dil08, Qin09, Zhang10}. The distance between impurity atoms is $a$. Rashba spin-orbit coupling in 2D electron system leads to a formation of the helical spin density wave in the 1D magnetic impurity chain.}
        \label{fig:schematic}
\end{figure}

Magnetism originating from interactions between magnetic atoms mediated by delocalized electrons (the so-called RKKY interaction) represents an important problem in modern condensed matter physics~\cite{RKKY} and has been a subject of intense research~\cite{Zutic2004, DS_PRB2003, Priour_PRL2006, Zhou10, Lobos12, Lobos13}. In this Letter, we consider the specific case of a helical spin density wave (SDW) that might appear in a one-dimensional chain of magnetic atoms that are coupled to a metal or a superconductor.
The issue of RKKY-induced magnetism has recently taken on a new and unexpected interesting perspective in the physics of non-Abelian Majorana bound states (Majoranas)~\cite{ReadGreen, TQCreview}, with the recent claims of the natural (i.e. self-tuned) emergence of Majorana modes in a chain of Yu-Shiba-Rusinov~\cite{Yu'65, Shiba68, Rusinov69} states induced by magnetic atoms at the surface of a superconductor, see Fig.~\ref{fig:schematic}. A Majorana-carrying topological superconducting phase should emerge in this system without the tuning of any external parameters due to the existence of an RKKY-stabilized helical order in conjunction with s-wave superconductivity~\cite{Yazdani'13, Loss'13, Simon'13, Franz'13}.  If correct, this is a breakthrough in the prospective realization of non-Abelian topological phases of matter, and hence of great importance. A helical spin texture is a crucial ingredient also in most other proposals for topological superconductivity~\cite{FuKane, Sau2010, Alicea2010, Lutchyn2010, Oreg2010, Sau2010a}.

In these recent Majorana proposals, the presence of the helical order was either assumed a priori~\cite{Choy2011, Kjaergaard2012, Martin2012, Nakosai2013, Yazdani'13, Pientka2013, Pientka2013b} or shown to exist in rather limited situations~\cite{Loss'13, Simon'13, Franz'13}. In this Letter, we revisit the claims of the emergent self-tuned topological superconductivity in magnetic chains in realistic experimental conditions. Specifically, we address the question whether the helical SDW in the perfectly ordered chain survives in the presence of disorder invariably present in physical systems. We consider two types of disorder: the positional disorder of the magnetic atoms and potential impurity scattering in the substrate. We find that the existence of the SDW necessary for creating Majoranas in the chain becomes severely constrained by disorder, and in fact, the SDW (and therefore, the topological superconductivity) is unlikely to emerge unless a strong spin-orbit (SO) interaction is present in the system.

Our main results are the following: In the case of an impurity chain coupled to a 1D conductor, we show that the helical SDW emerging due to the well-known $2k_F$ instability is stable provided that the fluctuations of the impurity positions are smaller than the Fermi wavelength of the underlying metallic substrate, which is an important constraint to satisfy since this means that the impurity atoms must form a periodic chain to better than a few angstroms precision. In the opposite regime, where the positional disorder becomes comparable to the Fermi wavelength, the impurity spins form ferro- or antiferromagnetic domains. In the case where the impurity chain is coupled to a 2D conductor (2DEG), see Fig.\ref{fig:schematic}, the effective range of the RKKY interaction becomes smaller and the helical SDW does {\it not} form spontaneously. Nevertheless, it is possible to stabilize the helical SDW if the 2D conductor breaks inversion symmetry, allowing Dzyaloshinskii-Moriya (D-M) interactions between the impurity spins. The latter favors a helical SDW with a pitch angle that depends on the relative strength of RKKY and D-M interactions. In order to study the robustness of the helical SDW in the 2D setup we consider a minimal model that generates D-M interaction, a 2D electron gas with Rashba SO coupling~\cite{Imamura2004}. We study the magnetic ordering in such a system in the presence of potential impurities in the 2DEG, and show that the helical SDW is stable in such a system provided the interatomic distance between impurity spins is smaller than the effective carrier mean-free path in the conductor. This result suggests that magnetic chains with D-M interactions coupled to a thin film superconductor (see Fig.\ref{fig:schematic}) can be used to create Majorana zero-energy states and study their properties.

We consider a 1D chain of magnetic impurity atoms coupled to 1D or 2D conduction electrons with Rashba interactions. The schematic picture of the experimental system is shown in Fig.~\ref{fig:schematic}. The corresponding effective Hamiltonian is given by ($\hbar=1$)
\begin{align}\label{eq:Hamiltonian}
H\!\!=\!\!\!\int\!\! d{\vec r}\!\left[c_{\alpha}^\dag \! \left(\frac{{\vec p}^2}{2m^*}\!-\!\mu\!+\! \alpha \hat{z}\!\cdot\!(\vec{p}\!\times\!\vec{\sigma})\!\right)_{\!\!\alpha \beta}\!\!\!\!c_{\beta}\!+\!J\vec{S}(\vec{r})\!\cdot\!\vec{s}(\vec{r})\!\right]
\end{align}
where $c_{\alpha}^\dag$ ($c_{\alpha}$) are the conduction electron creation (annihilation) operators with spin $\alpha$, $\sigma_i$ are Pauli matrices, $\vec{p}=-i \vec{\nabla}$ is the momentum operator; $\vec{S}$ and $\vec{s}$ are impurity and electron spin operators. $m^*$ is the effective mass of electrons and $J$ is the coupling strength. We assume that magnetic atoms such as $\rm Co$, $\rm Gd$ or $\rm Fe$ have large spin so that one can neglect quantum effects and treat the impurity spins as classical. 
By integrating out conduction electrons, one arrives at the following Hamiltonian for the impurity spins
\begin{equation}\label{eq:classical_H_simple}
H_\text{RKKY}=-J^2\sum_{ij}\sum_{\alpha,\beta}\chi_{\alpha\beta}(R_{ij})S_i^\alpha S_j^\beta.
\end{equation}
Here $\chi_{\alpha\beta}(R_{ij})$ is spin-spin susceptibility with $R_{ij} = |R_i - R_j|$ being the distance between two impurity spins. The real-space spin-spin correlation function $\chi_{\alpha\beta}(R)$ is given by
\begin{equation}
\chi_{\alpha\beta}(R)=-\int\frac{d \omega}{2\pi}\mathrm{Tr}\,[\sigma_\alpha G(\omega, R)\sigma_\beta G(\omega, -R)],
\end{equation}
where $G(\omega, R)$ is the Green's function for the conduction electrons.
\begin{figure}
\includegraphics[width=\columnwidth]{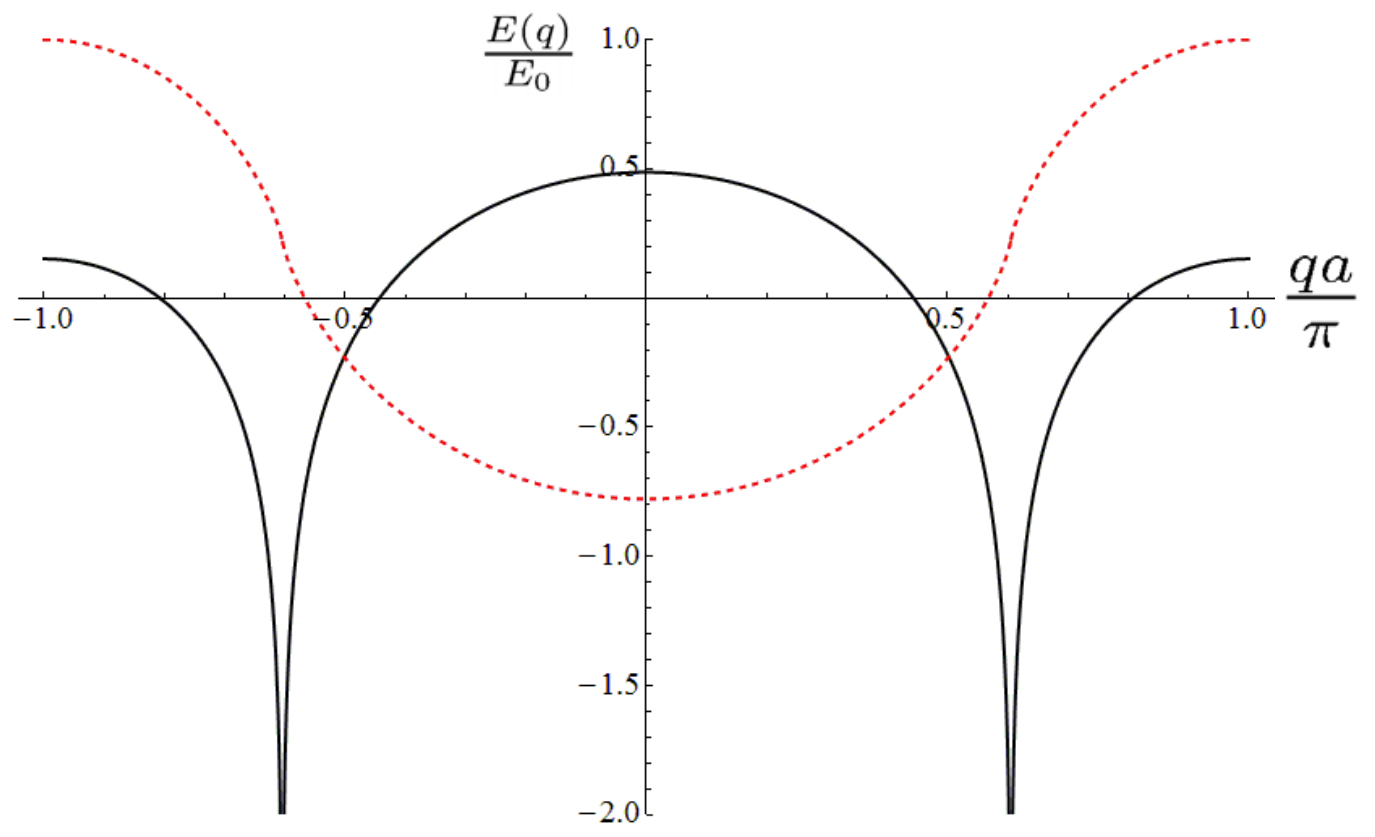}
        \caption{(Color online) The energy $E(q)$ vs variational parameter $q$ for 1D (solid black) and 2D (dashed red) RKKY interaction. $2k_F$ instability in 1D spin-spin susceptibility leads to the formation of helical spin density wave with $2k_Fa$ pitch angle. Here we used $2k_Fa = 1.9$. For 2D RKKY interaction, the minimum of the function $E(q)$ is at $q=0$, and, thus, the system favors ferromagnetic spin alignment.}
		\label{fig:1Dvs2D}
\end{figure}

We first consider the model without SO coupling $\alpha=0$, in which case the spin susceptibility is isotropic $\chi_{\alpha \beta}(R)\propto \delta_{\alpha \beta}F(k_F R)$. Here the range function $F(k_F R)$ describes RKKY interaction between impurity spins. The function $F(x)$ for 1D/2D conductor is well-known~\cite{Litvinov98}:
\begin{align}\label{eq:FR}
F_{\rm 1D}(x)\!&=\!-\left( {\rm Si}(2x)-\frac{\pi}{2}\right), \\
F_{\rm 2D}(x)\!&=\!-\frac{\pi}{4}\left[J_0(x)N_0(x)+J_1(x)N_1(x)\right]\nonumber.
\end{align}
Here ${\rm Si}(x)$ is the sine integral function, $J_0(x)$ and $N_0(x)$ are Bessel functions of the first and second kind, respectively, and $x=k_F R$ with $k_F$ being the Fermi momentum. In the limit $ x\gg 1$, $F_{\rm 1D}(x)\approx \frac{\cos(2x)}{2x}$ and $F_{\rm 2D}(x)\approx \frac{\sin(2x)}{(2x)^2}$.

We can make an ansatz for the ground state configuration where all spins lie in the same plane (for example the XY plane), but rotate by a pitch angle $q$ with respect to each other, $S_i = (\cos(q R_i), \sin(q R_i), 0)$; we will later show numerically that such an ansatz is justified. We thus have $S_i \cdot S_j = \cos(q |R_i-R_j|)$. In order to find the ground state spin configuration, one has to minimize the energy, $-\sum_{j=1}^\infty \cos(q a j)F(a j)$. In the limit $k_F a \gg 1$,
\begin{equation}\label{eq:sum}
\frac{E(q)}{E_0}=\!\begin{cases}
	\frac{1}{2}\!\left[\log\left|2 \sin\left[\frac{(2k_F\!-\!q)a}{2}\right]\right|+(q\rightarrow - q)\right]  \text{\,\, \,\,\,\,\,\,\,\,\,\,\,\,1D}\\
	\frac{i}{4}\!\left[{\rm Li}_2(e^{i(2k_F\!-\!q)a})+{\rm Li}_2(e^{i(2k_F\!+\!q)a})-\text{c.c.} \right] \text{\,\,\,2D}
\end{cases}
\end{equation}
with ${\rm Li}_2(x)$ being the polylogarithm function and $E_0=J^2 m^* k_F^2/2 \pi^2$. The plot of the functions $E_{1D/2D}(q)$ is shown in Fig.~\ref{fig:1Dvs2D}. One can notice that $E_{1D}(q)$ is sharply dipped at $q=\pm 2k_F$, which is simply a reflection of the long-range nature of the RKKY interaction in 1D, see Eq.~\eqref{eq:FR}, since the sum in Eq.~\eqref{eq:sum} diverges at these values. This is the reflection of $2k_F$ divergence in the spin susceptibility $\chi(q)$. On the other hand, the function $F_{\rm 2D}(x)$ decays much faster, the sum in Eq.~\eqref{eq:sum} is convergent for any value of $q$. As follows from Fig.~\ref{fig:1Dvs2D}, the spin configuration that minimizes the energy function corresponds to the ferromagnetic spin alignment. Thus, the spontaneous SDW formation~\cite{Loss'13, Simon'13, Franz'13} is a purely 1D effect, which critically relies on Fermi surface nesting. This observation rules out most of 2D or 3D metals and superconductors, where the Fermi surface does not generically have any nesting, for having $2k_F$ singularity. One interesting possibility is to consider a magnetic spin chain coupled to a quasi-1D superconductor with an open Fermi surface where tunneling matrix elements along the chain $t_{||}$ are much stronger than perpendicular to the chain $t_{\perp}$, i.e. $|t_{||}|\gg |t_{\perp}|$.

We now investigate the stability of the helical SDW in 1D structures against disorder in realistic experimental conditions, e. g., taking into account positional disorder of the magnetic impurities. The relevant length scale to which the positional disorder must be compared is the Fermi wavelength. This may be particularly short in a metal, typically few angstroms, potentially making this problem a crucial one. In this case even small deviations of the impurity positions from perfect periodicity may lead to frustration
of the magnetic interaction which would ultimately destroy the helical SDW. We study this question numerically using a simulated annealing procedure which allows us to identify the ground state spin configuration of the system in the presence of positional disorder. In our simulations, positional disorder is characterized by a scale $\delta R$, and atom positions are chosen as $R_i = a i + r$, where $r \in [-\delta R,\delta R]$ is uniformly chosen. We consider 50 disorder realizations for each parameter set, and for each disorder realization perform annealing with a local-update Monte Carlo procedure followed by a gradient-based energy optimization for 400 initial configurations. We analyze the lowest-energy configurations we obtain using the pitch angle of adjacent spins, $\theta_{i} = \arccos (\vec{S}_i \cdot \vec{S}_{i+1} )$. Furthermore, we confirm that all spins lie in a plane by calculating $T_{i} = (\vec{S}_i \times \vec{S}_{i+1}) \cdot \vec{S}_{i+2}$ and confirming that $\langle | T_{i} | \rangle = 0$ within error bars; this justifies the ansatz chosen above.

\begin{figure}
  \includegraphics[width=\columnwidth]{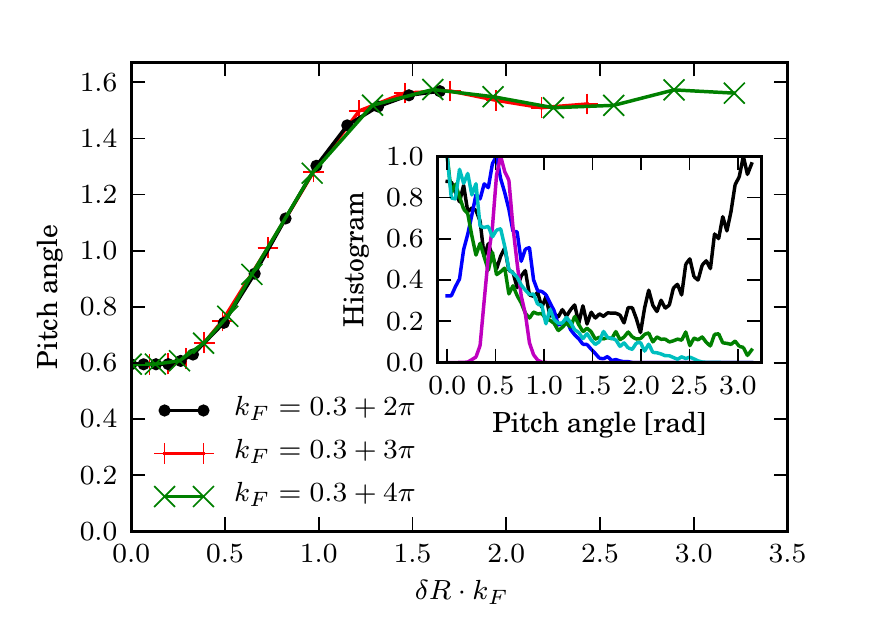}
  \caption{(Color online)
  {\it Main panel:} Average pitch angle vs $dR \cdot k_F$ for $F(R) = F_\text{1D}(R)$ of Eqn.~\eqnref{eq:FR} for $L=96$ sites.
  {\it Inset:} Histogram of the pitch angle for the same model as the main panel for $k_F \cdot \delta R = $ 0.07 (purple), 0.2 (blue), 0.33 (turqouise), 0.66 (green), 1.65 (black).\label{fig:disorder} }
\end{figure}

In the main panel of Fig.~\ref{fig:disorder}, we show the pitch angle averaged over position within each system and different disorder realizations for various values of $k_F$ and $\delta R$. 
Our results confirm that the effect of disorder is governed by the product $k_F \cdot \delta R$, as indicated by the collapse of all curves in the figure.

To understand in more detail how positional disorder affects the low-energy configurations, consider the histograms shown in the inset of Fig.~\ref{fig:disorder}. For $k_F \cdot \delta R \leq 0.2$, the histogram is sharply peaked around the pitch angle of the clean case. For larger values of $k_F \cdot \delta R$, the peak height is drastically reduced and the peak is broadened. Upon further increasing disorder, the peak is rapidly split into two peaks at $\theta=0$ and $\theta=\pi$, where the peak at $\theta=0$ is higher. This is indicative of configurations with ferromagnetic clusters separated by domain walls. As the size of these clusters is reduced, the peaks at $\theta = 0,\pi$ become more balanced and ultimately have the same height. This is reflected by the mean pitch angle approaching $\pi/2$. A key observation is that already for small values of the disorder $k_F \cdot \delta R \ll 1$, where the mean value is well away from $\pi/2$, the system is not in a helical phase but instead is composed of ferromagnetic clusters. This is different from the heuristic disorder model assumed in Ref.~\onlinecite{Choy2011}.

Our calculation for positional disorder establishes positional disorder to be a severe constraint restricting the spontaneous emergence of a helical SDW in the system of impurity spins coupled to a 1D metal.
Even assuming that one might be able to reduce disorder in the positions of the magnetic impurities through very careful sub-nm control of the impurity placement, there are further effects such as thermal fluctuations and potential disorder in the 1D metal itself which generically suppress helical magnetic ordering. Furthermore, all Majorana proposals ultimately require a sizeable coupling to a bulk s-wave superconductor. However, the interaction of the impurity spins with superconducting electrons in 2D or 3D favors an (anti-)ferromagnetic ground state and competes with helical ordering. Thus, we conclude that the observation of a spontaneously formed helical SDW in 1D systems is quite challenging and unlikely unless special care is taken in reducing all types of disorder in the system --  the self-tuned helical SDW is only possible in an ideal theoretical model, not in the laboratory where disorder is inherently present.

The experimental observation of helical ordering in $\rm Fe$ chains deposited on Ir (001) surface~\cite{experimentSDW} indicates that D-M interaction is necessary to stabilize the helical SDW. The microscopic origin of the D-M interaction is complex and often associated with the presence of SO coupling. Therefore, we now consider an impurity spin chain coupled to a 2D conductor with Rashba SO interaction $\alpha\neq 0$, see Eq.~\eqref{eq:Hamiltonian}. After integrating out the conduction electrons that mediate the interactions between impurity spins, we arrive at the anisotropic model for impurity spins~\cite{Imamura2004}
\begin{equation}
\begin{split}
&H\!=\!-\frac{2J^2m^* q_F^2}{ \pi^2} \!\sum_{i, j} F_{2D}(q_F R_{ij}) \left( \cos(2 k_R R_{ij}\!) \vec{S}_i\!\cdot\!\vec{S}_j \right.\\
&\!+\!\left.\sin(2 k_R R_{ij} ) \hat{y}\!\cdot\!(\vec{S}_i\!\times\!\vec{S}_j)\!+\! [1\!-\!\cos(2 k_R R_{ij})] S_i^y S_j^y \right).
\end{split}
\label{eqn:classical_H}
\end{equation}
Here the function $F_{2D}(R_{ij})$ is defined in Eq.~\eqref{eq:FR}; $q_F=\sqrt{k_F^2+k_R^2}$ with $k_R=m^*\alpha$. In order to understand ground state properties of the  Hamiltonian~\eqref{eqn:classical_H}, it is instructive to perform the following local transformation: $\tilde{S}_i^{x/z}=\cos (2k_R R_i) S_i^{x/z}\pm \sin (2k_R R_i) S_i^{z/x}$ and $\tilde{S}_i^y=S_i^y$ which is simply an $\mathbb{SO}(3)$ rotation around the $y$ axis by the angle $2k_R R_i$.
In the new rotated basis, the Hamiltonian~\eqref{eqn:classical_H} contains only the RKKY interaction. As argued above, the ground state of an impurity chain coupled to a 2D conductor is ferromagnetic. Thus, one can unwind the rotation to obtain the actual spin ordering. A simple calculation indicates that the ground states in this case corresponds to a helical SDW with a pitch angle $2k_R a$. This result should be contrasted with the spontaneous helical SDW with the pitch angle given by $2 k_F a$. Thus, strong SO coupling is essential for the helical RKKY Majorana proposals to be realized in magnetic impurity chains.

We now analyze the effect of potential disorder scattering, which is relevant for the RKKY Majorana proposals involving disordered superconductors. As previously mentioned, potential disorder scattering randomizes magnetic interactions and therefore affects ordering of magnetic atoms. Before considering the case with SO coupling, it is useful to first discuss the RKKY case which has been extensively studied in the literature~\cite{Zyuzin1986, Bulaevski1986, Bergmann1987, Lerner'1993, Galitski'2002, DS_PRB2003, Kaminski2004, Priour_PRL2006, Chesi2010, Abanin'11}. It is well-known that the disorder-averaged spin susceptibility $\overline{\chi}(r)$ decays exponentially with the decay length $l_c$. At large distances $r \gg l_c$, however, the susceptibility $\overline{\chi_{\alpha \beta}}(r)$ does not represent interactions between impurity spins in a given sample. Indeed, the fluctuations of the interaction are considerably larger than its typical value. Thus, one has to consider sample-specific interactions which decay much slower than $\overline{\chi}(r)$, i.e. as a power law. As shown below, the situation is qualitatively similar in disordered metals with Rashba interactions. Therefore, at small distances between magnetic atoms ($a\ll l_c$) the short range nature of spin-spin interactions dominates and the system forms a helical spin density wave, similar to the clean case, whereas at large distances between magnetic impurity atoms ($a\gg l_c$) random RKKY interactions cause frustration and destroy magnetic order. In this sense, substrate random disorder scattering is similar to the positional disorder in the magnetic chain itself which drives the system into a paramagnetic phase, see Fig.~\ref{fig:disorder}.
\begin{figure}
\includegraphics[width=4.2cm]{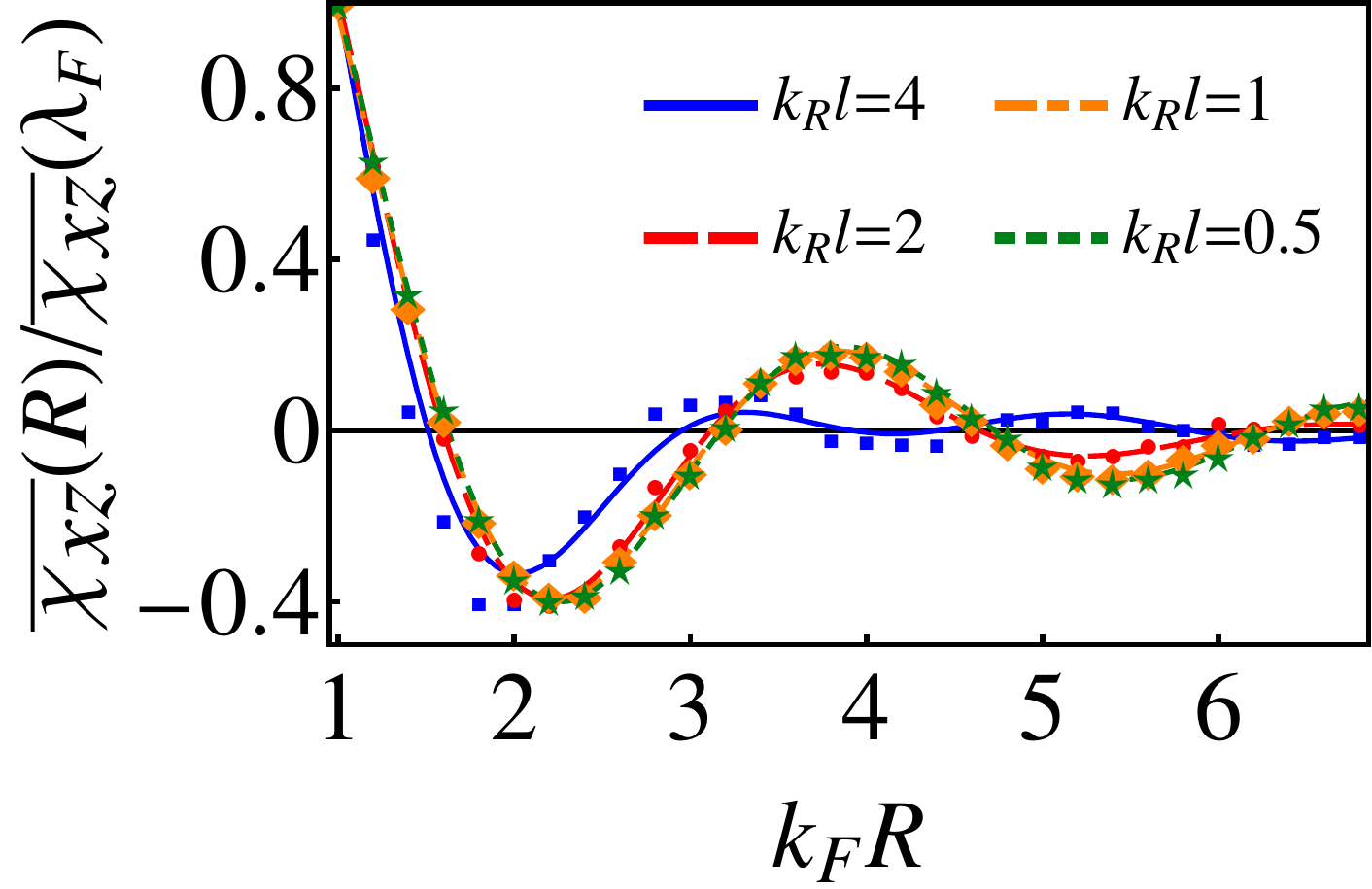}\,\,\includegraphics[width=4.2cm]{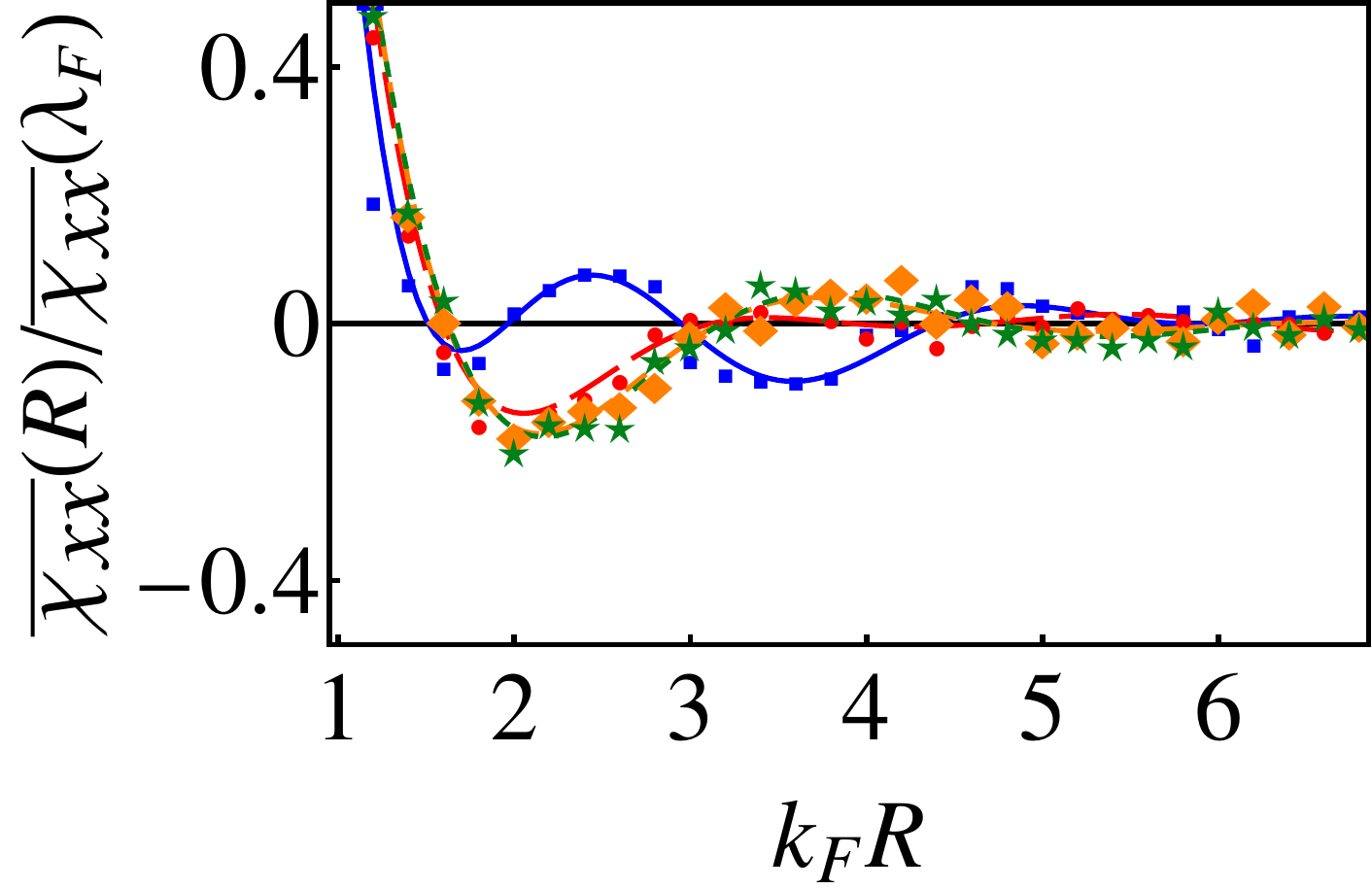}
\includegraphics[width=8.4cm]{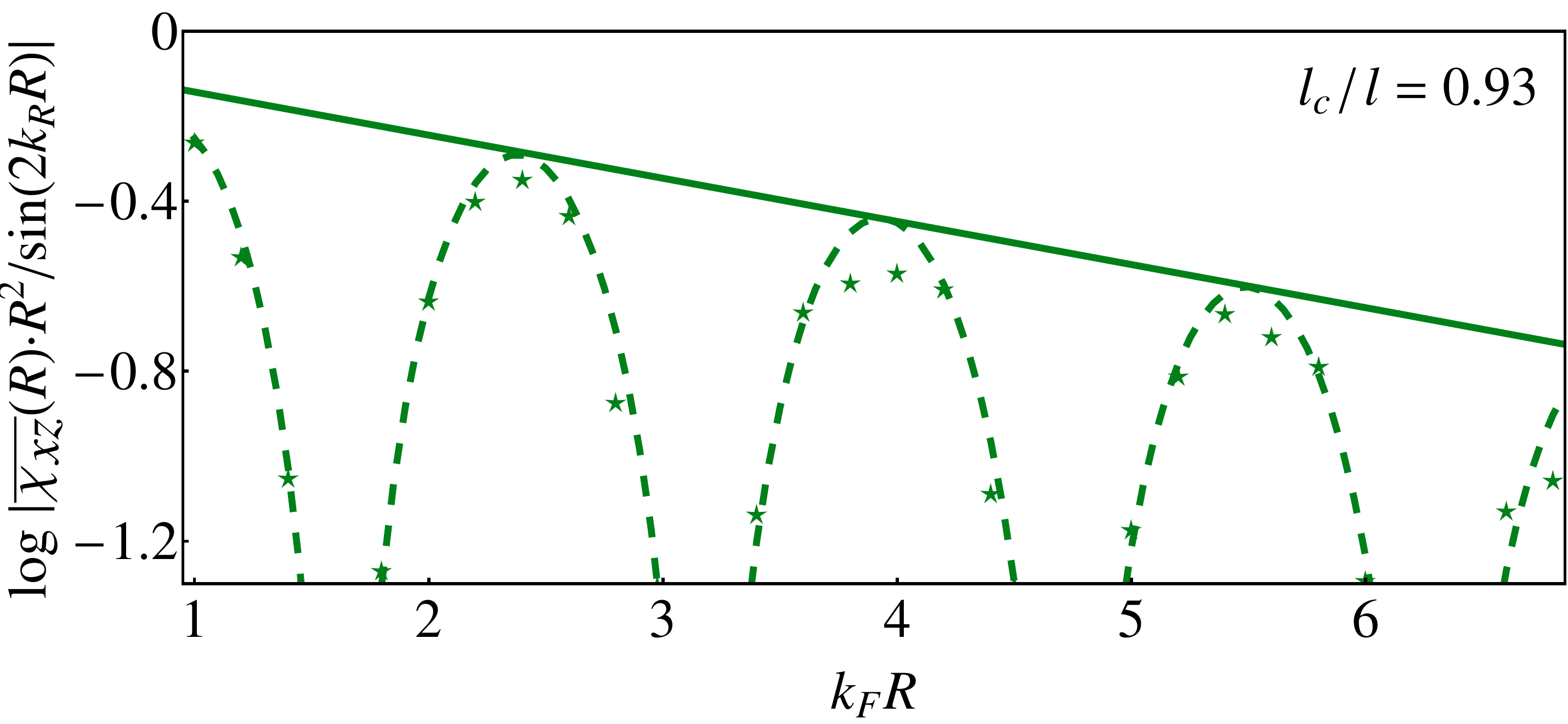}
\caption{Spatial dependance of the spin susceptibilities $\overline{\chi^0_{xx}}(R)$ and $\overline{\chi^0_{xz}}(R)$ in the presence of Rashba SO coupling. Here different colors correspond to different values of the SO interaction for a fixed $k_F l =10$, see legend in the top left panel. Both functions have spatially oscillating prefactor and exponentially decaying envelop. The fit of the exponential decay allows one to extract $l_c$, see the bottom panel.}
\label{fig:suscept}
\end{figure}

In order to estimate the characteristic decay length $l_c$, we calculate the spin-spin susceptibility tensor $\overline{\chi_{\alpha \beta}}(r)$ in the presence of SO coupling. We employ a standard disorder diagrammatic technique, see Supplementary Material for details. We consider a model with Gaussian random disorder, where the disorder potential $V(\vec{r})$ is $\delta$-correlated: $\langle V(\vec{r})V(\vec{r}')\rangle=(2\pi \nu_F \tau)^{-1}\delta(\vec{r}-\vec{r}')$ where $\nu_F$ and $\tau$ are the density of states at the Fermi level and impurity scattering time, respectively. Our main results are summarized in Fig.~\ref{fig:suscept}. Since SO coupling breaks the $\mathbb{SU}(2)$ symmetry, we now have four non-zero spin-spin correlation functions $\overline{\chi_{xx}}(q)$, $\overline{\chi_{yy}}(q)$, $\overline{\chi_{zz}}(q)$  and $\overline{\chi_{xz}}(q)$. Their spatial dependence is characterized by an oscillatory pre-factor and an exponentially-decaying envelope function, see Fig.~\ref{fig:suscept}. The prefactor has a spatial dependence which is very similar to the spin-spin interaction in the clean limit, see Eq.~\eqref{eqn:classical_H}. The characteristic decay length $l_c$ can be obtained by fitting the envelope function. In the limit of $k_R \ll k_F$, $l_c$ is very weakly dependent on the SO interaction strength, and is determined by the mean-free path $l$ (up to a numerical prefactor of order one).

Having established the limitations on the stability of the helical SDW, we now discuss the helical RKKY Majorana proposals~\cite{Yazdani'13, Loss'13, Simon'13, Franz'13, Pientka2013} and compare them with the semiconductor nanowire ones~\cite{Lutchyn2010, Oreg2010} which have recently been studied extensively experimentally~\cite{Mourik2012, Das2012,Deng2012,  Churchill2013, Finck_PRL2013, Rokhinson2012a}. As shown above, spontaneous formation of the helical SDW with a pitch angle $2k_Fa$ critically relies on one-dimensionality. One of the systems that has been put forward involves nuclear spins coupled to 1D semiconductor electrons~\cite{Loss'13, Simon'13}. However, the crossover temperature $T^*$ above which helical order disappears is very low in this system ($T^*\!\sim\! 1$mK) due to the small hyperfine coupling. When coupled to a higher-dimensional conductor, the ground state magnetic ordering is ferromagnetic in the absence of D-M interaction. Therefore, in realistic experimental situations a large SO coupling is necessary for the realization of the helical SDW. In this case, the pitch angle of the helical order is set by the SO wave length. The {\it only} evidence for helical order in chains of magnetic atoms comes from the experiment~\cite{experimentSDW} involving Fe atoms placed on an Ir(001) surface. This supports our conclusion, since the D-M interaction is very large in this experiment.

Using a thin film superconductor with strong SO coupling one can realize Majorana bound states in the magnetic atom chains on top of it. A particularly promising system involves an ultrathin Pb film deposited on Si$(111)$ substrate which has the superconducting transition temperature $T_c\approx 6$K~\cite{Qin09, Zhang10} as well as a large Rashba SO coupling of $k_R=0.035\AA^{-1}$~\cite{Dil08}. Also, one can grow atomically uniform Pb films on Si(111) surface\cite{Upton04}, so the mean free path is expected to be much larger than the interatomic spacing. Thus, this system satisfies the minimal conditions necessary to realize topological superconductivity. However, the pitch angle is now determined by the SO coupling strength rather than the ``sweet spot" value $2k_Fa$.  Therefore, some tuning is generically necessary in order to drive the system into the topological superconducting phase~\cite{Yazdani'13, Pientka2013, Pientka2013b}, which might be quite challenging in the aforementioned setup. This is to be contrasted with the semiconductor-based Majorana proposals~\cite{Sau2010, Alicea2010, Lutchyn2010, Oreg2010} where one can tune the magnetic field to drive the system into a topological superconducting phase. We emphasized the role of random disorder scattering on the stability of the helical order; however, it also has a detrimental effect on the stability of the topological superconducting state~\cite{Motrunich01, Gruzberg05, Brouwer11, LobosPRL'12}. Our finding that magnetic impurities form ferromagnetic domains has implications for helical RKKY Majorana proposals since such ordering affects proximity-induced superconducting pairing and therefore suppresses the topological phase. On the positive side, we believe that a big advantage of the RKKY Majorana proposal is the ability to detect zero-energy bound states directly using STM rather than tunneling transport measurements as suggested originally in Ref.~\cite{Sau2010a}.


\acknowledgements
This work is supported by Microsoft Q and JQI-NSF-PFC. We thank A. Bernevig, L. Glazman and A. Yazdani for discussions.
Simulations were performed using the ALPS libraries~\cite{bauer2011-alps}.

\bibliography{RKKYRef}
\end{document}